# Photon-counting optical coherence-domain reflectometry using superconducting single-photon detectors

**Nishant Mohan[1*], Olga Minaeva[2,3], Gregory N. Gol'tsman[3], Magued B. Nasr[2], Bahaa E. A. Saleh[2], Alexander V. Sergienko[2,4], and Malvin C. Teich[1,2,4]**

[1]*Department of Biomedical Engineering, Boston University, Boston, MA 02215*
[2]*Department of Electrical and Computer Engineering, Boston University, Boston, MA 02215*
[3] *Department of Physics, Moscow State Pedagogical University, Moscow 119992, Russia*
[4]*Department of Physics, Boston University, Boston, MA 02215*
[*]*Corresponding author:  nm82@bu.edu*

**Abstract:** We consider the use of single-photon counting detectors in coherence-domain imaging. Detectors operated in this mode exhibit reduced noise, which leads to increased sensitivity for weak light sources and weakly reflecting samples. In particular, we experimentally demonstrate the possibility of using superconducting single-photon detectors (SSPDs) for optical coherence-domain reflectometry (OCDR). These detectors are sensitive over the full spectral range that is useful for carrying out such imaging in biological samples. With counting rates as high as 100 MHz, SSPDs also offer a high rate of data acquisition if the light flux is sufficient.

©2008 Optical Society of America

**OCIS codes:** (110.4500) Optical coherence tomography; (040.5160) Photodetectors; (030.5260) Photon counting

**References and links**

1. A. F. Fercher, W. Drexler, C. K. Hitzenberger, and T Lasser, "Optical coherence tomography—principles and applications," Rep. Prog. Phys. **66**, 239-303 (2003).
2. S. A. Boppart, B. E. Bouma, C. Pitris, J. F. Southern, M. E. Brezinski, and J. G. Fujimoto, "*In vivo* cellular optical coherence tomography imaging," Nature Med. **4,** 861-865 (1998).
3. W. Drexler, "Ultra-high resolution optical coherence tomography," J. Biomed. Opt. **9**, 47-74 (2004).
4. I. Hartl, X. D. Li, C. Chudoba, R. K. Ghanta, T. H. Ko, and J. G. Fujimoto, "Ultrahigh-resolution optical coherence tomography using continuum generation in an air–silica microstructure optical fiber," Opt. Lett. **26**, 608-610 (2001).
5. B. Povazay, K. Bizheva, A. Unterhuber, B. Hermann, H. Sattmann, A. F. Fercher, W. Drexler, A. Apolonski, W. J. Wadsworth, J. C. Knight, P. St. J. Russell, M. Vetterlein, and E. Scherzer, "Submicrometer axial resolution optical coherence tomography," Opt. Lett. **27**, 1800-1802 (2002).
6. S. Carrasco, M. B. Nasr, A. V. Sergienko, B. E. A. Saleh, M. C. Teich, J. P. Torres, and L. Torner, "Broadband light generation by noncollinear parametric downconversion," *Opt. Lett.* **31**, 253-255 (2006), co-published in *Virtual Journal of Biomedical Optics*.
7. B. E. A. Saleh and M. C. Teich, *Fundamentals of Photonics*, 2nd Ed. (Wiley, 2007), Chaps. 11, 12, 18, and 24.
8. M. C. Teich, "Field-theoretical treatment of photomixing," *Appl. Phys. Lett.* **14**, 201-203 (1969).
9. S. Carrasco, M. B. Nasr, A. V. Sergienko, B. E. A. Saleh, M. C. Teich, J. P. Torres, and L. Torner, "Broadband light generation by noncollinear parameteric downconversion," Opt. Lett. **31**, 253-255 (2006).
10. A. F. Fercher, C. K. Hitzenberger, M. Sticker, E. Moreno-Barriuso, R. Leitbeg, W. Drexler, and H. Sattmann, "A thermal light source technique for optical coherence tomography," Opt. Comm. **185**, 57-64 (2000).
11. Y. Wang, Y. Zhao, J. S. Nelson, and Z. Chen, "Ultrahigh-resolution optical coherence tomography by broadband continuum generation from a photonic crystal fiber," Opt. Lett. **28**, 182-184 (2003).
12. G. N. Gol'tsman, K. Smirnov, P. Kouminov, B. Voronov, N. Kaurova, V. Drakinsky, J. Zhang, A. Verevkin, and R. Sobolewski, "Fabrication of nanostructured superconducting single-photon detectors," IEEE Trans. Appl. Supercond. **13**, 192-195 (2003).

13. G. N. Gol'tsman, O. Okunev, G. Chulkova, A. Lipatov, A. Semenov, K. Smirnov, B. Voronov, and A. Dzardanov, "Picosecond superconducting single-photon optical detector," Appl. Phys. Lett. **79**, 705-707 (2001).
14. G. N. Gol'tsman, A. Korneev, I. Rubtsova, I. Milostnaya, G. Chulkova, O. Minaeva, K. Smirnov, B. Voronov, W. Słysz, A. Pearlman, A. Verevkin, and R. Sobolewski, "Ultrafast superconducting single-photon detectors for near-infrared-wavelength quantum communications," Phys. Status Solidi (c) **2**, 1480-1488 (2005).
15. M. E. Brezinski, *Optical Coherence Tomography: Principles and Applications* (Academic, 2006).
16. W. V. Sorin and D. M. Baney, "A simple intensity noise reduction technique for optical low-coherence reflectometry," IEEE Photon. Tech. Lett. **4**, 1404-1406 (1992).
17. A. G. Podoleanu, "Unbalanced versus balanced operation in an optical coherence tomography system," Appl. Opt. **39**, 173-182 (2000).
18. B. E. Bouma and G. J. Tearney, "Power-efficient nonreciprocal interferometer and linear-scanning fiber-optic catheter for optical coherence tomography," Opt. Lett. **24**, 531-533 (1999).
19. M. C. Teich, "Infrared Heterodyne Detection," *Proc. IEEE* **56**, 37-46 (1968).
20. M. C. Teich, "Quantum Theory of Heterodyne Detection," in *Proc. Third Photocond. Conf.*, edited by E. M. Pell (Pergamon Press, New York, 1971), pp. 1-5.
21. M. C. Teich and B. E. A. Saleh, "Photon Bunching and Antibunching," in *Progress in Optics*, vol. **26**, edited by E. Wolf (North-Holland/Elsevier, Amsterdam, 1988), ch. 1, pp. 1-104.
22. S. B. Lowen and M. C. Teich, *Fractal-Based Point Processes* (Wiley, 2005), Chap. 3.
23. T. S. Larchuk, M. C. Teich, and B. E. A. Saleh, "Statistics of Entangled-Photon Coincidences in Parametric Downconversion," *Ann. N. Y. Acad. Sci.* **755**, 680-686 (1995).
24. H. Lim, Y. Jiang, Y. Wang, Y. Huang, Z. Chen, and F. W. Wise, "Ultrahigh-resolution optical coherence tomography with a fiber laser source at 1 $\mu$m," Opt. Lett. **30**, 1171-1173 (2005).

## 1. Introduction

Over the past decade, optical coherence-domain techniques such as optical coherence-domain reflectometry (OCDR) and optical coherence tomography (OCT) have come into their own for use in biological imaging [1,2]. These techniques operate on interferometric principles and use heterodyne detection to achieve high detection sensitivity. In scattering tissue, they typically provide axial resolution of a few micrometers and imaging at depths of 2–3 millimeters.

The central wavelength of the light used in coherence-domain imaging is a key parameter of the system design. Optical scattering in biological tissue generally decreases with increasing wavelength. It is usually difficult to image deeply into tissue in the visible region so that most coherence-domain imaging systems make use of light sources with wavelengths longer than 700 nm. The long-wavelength limitation is governed by the absorption of water, which becomes problematical at about 1500 nm. Since the axial resolution of a coherence-domain imaging system improves as the spectral bandwidth of the light source increases, use of the entire wavelength range from 700 to 1500 nm yields a desirable combination of deep penetration and ultra-high resolution for biological tissue. Thus, broadband operation at a center wavelength near 1100 nm is advantageous for ultra-high-resolution coherence-domain imaging, assuming that there is a suitable detector in this region [3].

A number of high-axial-resolution coherence-domain imaging experiments using ultra-broadband light sources have indeed been reported over the past few years. However, because of the ready availability of commercial semiconductor photodetectors that operate near 800 nm and 1300 nm, most of these systems have been operated near one of these two wavelengths [4,5,6].

In this paper, we report the development of a photon-counting optical coherence-domain imaging system that makes use of superconducting single-photon detectors (SSPDs). Such detectors are sensitive over a broad wavelength band, including the region of interest for biological imaging, thus allowing for flexibility in the choice of operating wavelength. At the same time, they operate in a single-photon counting mode, which offers low detector noise and thereby provides high sensitivity even at low source powers.

## 2. Conventional OCDR

As indicated above, the high detection sensitivity of coherence-domain imaging results from the use of heterodyne detection. As illustrated in Fig. 1, the interference signal that results from the mixing of light from the reference and sample arms carries the information of interest. The magnitude of the interference signal is proportional to the product of the optical fields reflected from the two arms of the interferometer, and thus to the square-root of the product of the intensities reflected from these arms. The strong reference beam provides conversion gain, which effectively boosts the weak signal reflected from the sample [7]. It has been shown that the heterodyne process can be understood in terms of the absorption of individual polychromatic photons [8].

Conventional optical sources used in coherence-domain imaging usually provide sufficient power in the reference beam to achieve shot-noise limited operation with ordinary photodiodes. However, some optical sources with large bandwidths and smooth spectra [9,10], which are particularly useful for coherence-domain techniques, do not provide sufficient power in a single spatial mode to allow shot-noise-limited operation.

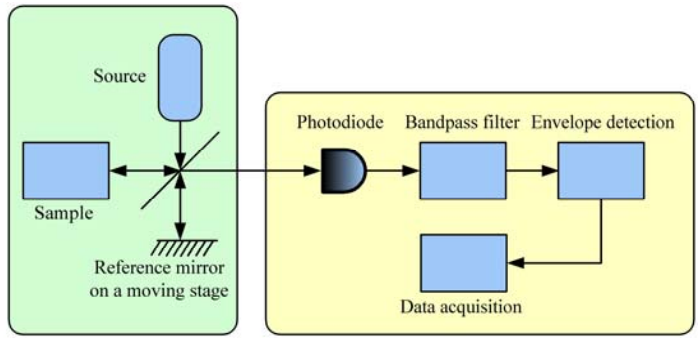

Fig. 1. Schematic of a conventional coherence-domain reflectometry (OCDR) experiment.

For the most part, OCDR and OCT experiments make use of commercially available Si or InGaAs semiconductor photodiodes (operated without gain), depending on the spectrum of the light source employed. Roughly speaking, Si photodiodes are used for wavelengths shorter than 1100 nm and are best in the vicinity of 800 nm, whereas InGaAs photodiodes are used for wavelengths longer than 1100 nm and are designed for operation in the vicinity of 1300 nm. Inasmuch as neither Si nor InGaAs are sensitive over the entire spectral range useful for the imaging of scattering biological samples, ultra-high-resolution OCDR and OCT is usually carried out at a central wavelength of either 800 nm or 1300 nm.

Comparing coherence-domain imaging at 800 nm and 1300 nm, we recognize that the latter wavelength offers superior penetration depth but inferior axial resolution. This is because the axial resolution, for a given spectral bandwidth specified in terms of wavelength, is inversely related to the square of the central wavelength. However, an ultra-broadband source of light centered at 1100 nm can provide the best of both worlds: deep penetration together with high resolution. This has indeed been demonstrated by Wang *et al.* [11], who achieved a resolution of 1.8 $\mu$m at a wavelength of 1100 nm. The performance of their system was limited, however, by the insensitivity of their detector to the shorter wavelength portion of their source spectrum.

As the use of ultra-broadband spectra in biological coherence-domain imaging becomes more widespread, there is a growing need for sensitive detectors that can operate over the entire wavelength range of interest to jointly optimize both axial resolution and penetration depth.

### 3. Photon-Counting OCDR

We have carried out a series of experiments to demonstrate the merits of using SSPDs in OCDR. These detectors are sensitive over a broad range of wavelengths, making them a good candidate for use in high-resolution coherence-domain techniques that require a broad spectrum of light. Moreover, since SSPDs operate in a photon-counting mode, they also offer enhanced sensitivity for low levels of light. We discuss the photon-counting OCDR system configuration, and the operational principles and properties of SSPDs, in turn.

*3.1 Experimental arrangement for photon-counting-based OCDR*

The photon-counting OCDR system illustrated in Fig. 2 makes use of the same interferometric arrangement as employed in standard coherence-domain imaging (Fig. 1). The reference arm of the interferometer has a mirror placed on a scanning delay stage, which is controlled by a Nanomotion-II micropositioning system (Applied Precision, LLC, Issaquah, WA). The sample arm contains the sample under investigation. The light exiting from the interferometer is coupled to a single-mode fiber that feeds the SSPD. An incident photon causes the detector to generate an electrical pulse; the probability of such an occurrence depends on the quantum efficiency of the detector. Once produced, the pulse is amplified and fed to a discriminator, which generates a standardized electrical pulse if the magnitude of the detector pulse lies above a prespecified threshold. The output of the discriminator is processed by a PC using National Instrument's Data-Acquisition Counter-Timer (Model PCI 6602).

To obtain the axial profile of the sample of interest, the discriminator output is recorded as the reference mirror is continuously scanned. The numbers of pulses obtained in a user-defined counting time are assigned to the corresponding position of the reference arm. An alternate way of obtaining the axial profile is to move the reference mirror in discrete steps and to integrate the pulse count from the discriminator for a finite amount of time at each location. In both cases the discrete signal is then bandpass filtered and demodulated to obtain its envelope. The scanning, data acquisition, and synchronization are all performed in an automated fashion using LabView.

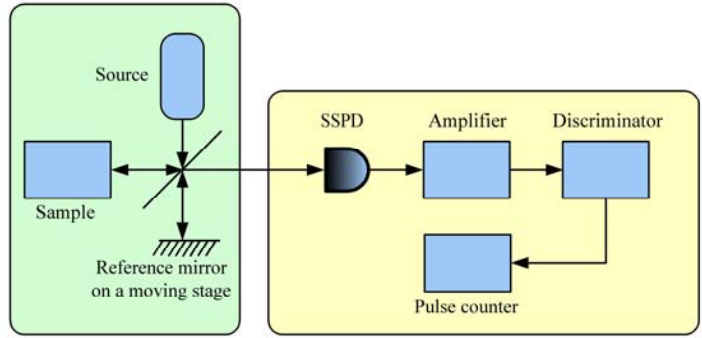

Fig. 2. Schematic of a photon-counting-based optical coherence-domain reflectometry (OCDR) experiment.

*3.2 Superconducting single-photon detectors*

The active element of the SSPD is a meander-shaped narrow stripe that covers the 10 $\mu$m x 10 $\mu$m area of the device. The stripe is fabricated from a 4-nm-thick superconducting niobium nitride (NbN) film that has been sputtered on a double-sided polished sapphire substrate, using direct electron-beam lithography and reactive ion etching [12]. The width of the stripe is 80-120 nm.

The SSPD operates by utilizing a resistive region that appears in the superconducting stripe following the absorption of a photon. This absorption creates a hotspot (a localized region with increased resistivity) that suppresses the superconductivity. The device is maintained at a temperature $T$ that is substantially below the critical temperature $T_c$. The device is electrically biased along its length by a current $I_b$ that is close to the critical current $I_c$. During the thermalization stage, the hotspot grows in size as electrons diffuse out of the initial hotspot core. The supercurrent is expelled from the hotspot into the side regions where its density exceeds the critical current density, thereby initiating the appearance of a resistive barrier across the entire cross-section of the stripe. This gives rise to a voltage pulse with a magnitude proportional to the bias current.

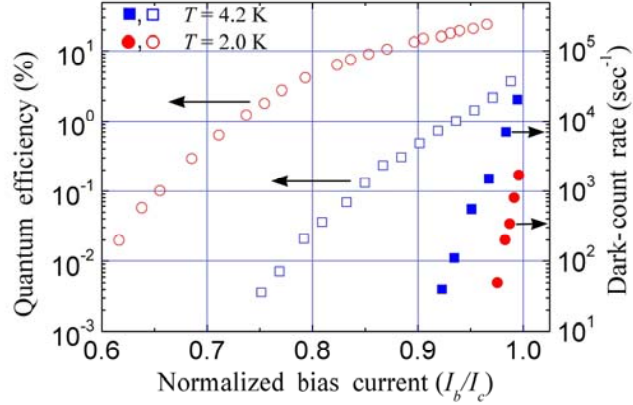

Fig. 3. Quantum efficiency and dark-count rate vs. normalized bias current at 1.3 $\mu$m for two different temperatures (4.2 K and 2.0 K).

Superconducting devices are very attractive for single-photon-detection applications, especially in the infrared region, because of their small energy gap $\Delta$ ($\Delta \approx 2$ meV for NbN) and their low dark-count rate.

The quantum efficiency $\eta$, defined as the probability of obtaining a voltage pulse at the SSPD output in response to an input photon, as well as the dark-count rate, strongly depend on the bias current and on the temperature of operation, as illustrated in Fig. 3 for light at a wavelength of 1.3 $\mu$m (the quantum efficiency in the figure is indicated in %). It is apparent that higher sensitivity and lower dark-count rate are achievable as the temperature is decreased.

The quantum efficiency of SSPDs monotonically decreases with increasing wavelength of the incident light. Despite this, these detectors can be reliably used for single-photon-counting applications in a spectral region that stretches from 0.4 to 6 $\mu$m [13]. Some semiconductor-based photodetectors can also serve as single-photon detectors in the infrared, but they suffer from a more limited wavelength range and from far higher dark-count rates.

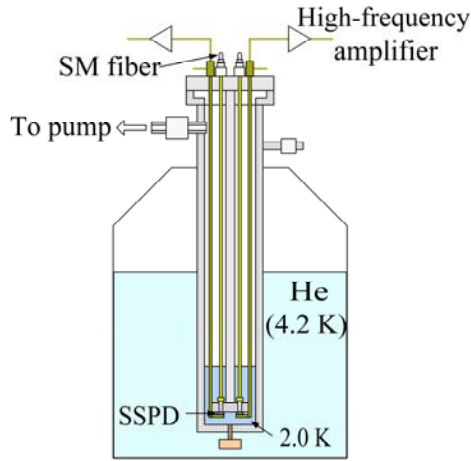

Fig. 4. A schematic of the system for low-temperature operation with an SSPD. Only one of the two SSPD channels is used in the current experiment.

Although SSPDs have attractive parameters for infrared single-photon counting, their use in practice is complicated by the need for low-temperature operation and by their small active area. To accommodate these requirements, we made use of a specially designed cryostat, outfitted with a superconducting detector fed by a single-mode (SM) fiber, as illustrated in Fig. 4. This allowed us to work efficiently with 10 $\mu$m x 10 $\mu$m detectors at selected temperatures ranging from 1.8 K to 4.2 K.

The input to the single-mode optical fiber is equipped with a standard FC connector, permitting use with various optical systems. The output of the detector is connected to a high-frequency coaxial cable through a coplanar RF transmission line. The apparatus is positioned inside a standard 60-liter liquid-helium transport dewar and the detectors can be cooled to 1.8 K by reducing the He vapor pressure. The room-temperature high-frequency amplifiers (Phillips Scientific 6954 0.0001-1.5 GHz) boost the electrical signals before they are fed to discrimination and counting circuitry.

Another advantage of the SSPD is its ability to carry out photon counting at repetition rates in excess of 100 MHz [14], which is large in comparison with many single-photon detectors. The oscilloscope-screen image portrayed in Fig. 5 shows that the SSPD response follows an incident train of light pulses presented at an 81.3-MHz repetition rate.

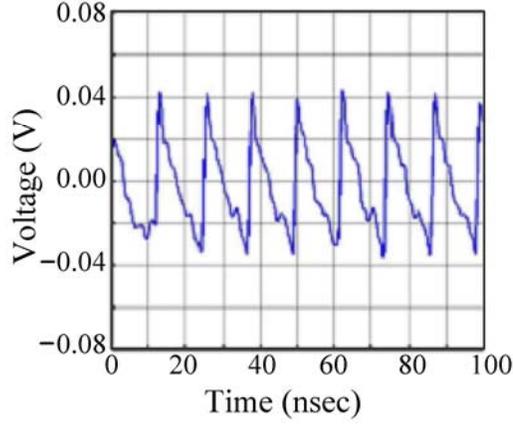

Fig. 5. Oscilloscope-screen image showing the response of an SSPD to an incident train of light pulses at an 81.3 MHz repetition rate.

## 4. Theory

*4.1 Axial resolution*

The axial resolution in coherence-domain imaging systems is governed by the bandwidth of the source, as well as the frequency response of the optical components and the detector. High axial resolution is attained by making use of a broadband source together with optical components and a detector that exhibit flat responses over the spectral range of interest. The usual array of optical components in use in such systems do indeed have approximately flat responses. The overall spectral response of the system, $S(n)$, is therefore given by $S(n) = S_S(n) \times S_D(n)$, where $S_S(n)$ is the spectrum of the source and $S_D(n)$ is the frequency response of the detector.

The point-spread function $f(\cdot)$ in coherence-domain imaging is proportional to the Fourier transform of the overall spectral response $S(n)$, so that [7,15]

$$f(2z/c) \propto \mathrm{FT}\{S_S(\nu) \cdot S_D(\nu)\}, \tag{1}$$

where $z$ is the reference-arm displacement in the interferometer, $c$ is the speed of light in the medium under consideration, and FT indicates the Fourier transform. The width of the point-spread function is the axial resolution $\Delta z$.

Of principal interest in this paper is the effect of the spectral response of the detector on axial resolution. Since the point-spread function is the convolution of the temporal coherence function of the source with the Fourier transform of the detector spectral response, the axial point-spread function will, by necessity, be wider than the coherence function of the source. A detector with a relatively flat and smooth spectral response function over the bandwidth of interest is best suited for coherence-domain imaging because it offers the least amount of broadening of the point-spread function.

*4.2 Sensitivity*

An oft-used measure for characterizing sensitivity is the signal-to-noise ratio SNR, where the signal is proportional to the optical power from the sample arm and the noise is defined as the variance of the background. Three principal sources of noise are generally considered: thermal electrical noise in the detector and post-detection circuitry, electric-current shot noise, and intensity-fluctuation noise arising from the thermal character of the optical source [7]. Noise-in-signal contributions are ignored in this definition.

An expression for the current SNR in standard time-domain OCDR and OCT experiments can be written as [1,16,17]

$$\text{SNR} = \frac{R^2 P_R P_S}{\frac{4kTB}{R_f} + 2eBRP_R + \frac{(1+\Pi^2)}{\Delta \nu} BR^2 P_R^2}, \quad (2)$$

where $P_R$ and $P_S$ are the optical powers in the reference and sample arms of the interferometer, respectively, and $R$ is the responsivity of the detector (A/W). The first term in the denominator represents the thermal noise in the receiver, where $T$ is the temperature, $k$ is Boltzmann's constant, $B$ is the effective electrical bandwidth of the detection system (which is principally determined by the bandpass filter following the detector), and $R_f$ is the feedback resistance of the trans-impedance amplifier. The second term in the denominator represents the current shot noise, where $e$ is the charge of an electron. The third term represents gamma-distributed intensity-fluctuation noise associated with the thermal nature of the light source; $\Pi$ is the degree of polarization of the light, and $\Delta\nu$ represents the spectral bandwidth of the light source [7].

The intensity-fluctuation noise term that depends on the square of the reference-beam optical power dominates at high values of $P_R$, whereas the detector thermal-noise term dominates at low values. Coherence-domain imaging systems typically operate at intermediate values of the reference-beam power, where shot noise is important [16,18], in which case Eq. (3) reduces to

$$\text{SNR} = \frac{RP_S}{2eB}. \quad (3)$$

Operation in this domain is considered desirable since it offers the largest signal-to-noise ratio for a given optical power in the sample arm.

The presence of detector thermal noise is sometimes unavoidable, however, if the light source cannot provide sufficient power to the reference arm. Taking the parameter values used by Soren and Baney [16], for example, using standard photodiode-based detection, detector thermal noise becomes significant for reference powers below 10 nW. However, it is important to observe that there is a way of reducing the contribution of detector noise by several orders of magnitude, so that it becomes insignificant even for pW levels of reference-beam optical power: use single-photon counting.

In photon-counting-based coherence-domain imaging, we record the number of photons at the output of the discriminator (see Fig. 2) in a given counting time of duration $T$; the corresponding bandwidth at the output of the photon-counting detector is $1/2T$ [7]. A single interferometric scan comprises a sequence of these counts collected at different positions of the reference-arm mirror. This sequence can be digitally filtered by using a bandpass filter with the same bandwidth as the signal, thereby reducing the noise. The bandwidth $B$ of the filtered interferometric scan is then that of the filter. It should be noted that digital bandpass filtering in photon-counting plays the same role as such filtering in conventional OCT (OCDR).

The relevant signal-to-noise ratio in the shot-noise regime is [7]

$$\text{SNR} = \frac{\eta \Phi_S}{2B}, \quad (4)$$

where $\Phi_S$ is the photon flux from the sample arm (photons arriving at the detector per sec). At an SNR of unity, it is apparent that the minimum-detectable photon flux is given by

$$\Phi_S^{\min} = \frac{2B}{\eta}. \tag{5}$$

This signifies the detection of $1/\eta$ photons per resolution time of the receiver, which, for unity quantum efficiency, corresponds to the detection of one photon per resolution time, which is optimal [19,20].

In addition to the signal-to-noise ratio, we can also consider the statistical nature of the photon counts of the signal. These fluctuations can be evaluated by determining the ratio of count-variance to count-mean [21,22],

$$F = \frac{\text{var}(n)}{\langle n \rangle}. \tag{6}$$

This quantity is also known as the normalized variance or the Fano factor [20]. For independent measurements at a given mirror location, and a source that is devoid of intensity fluctuations, we expect the counts to follow Poisson statistics. The Poisson distribution has mean $\langle n \rangle$ and variance $\text{var}(n) = \langle n \rangle$, so that $F = 1$. In real measurements, however, we have a finite number of samples $N$, and can therefore only obtain an estimate of the normalized variance $F$. This estimate, which we denote $\hat{F}$, is itself a random variable with a mean of unity and a standard deviation that turns out to be $\sqrt{2/N}$ for Poisson statistics [23].

*4.3 Data acquisition rate*

The rate of acquiring data in conventional coherence-domain imaging is rarely limited by the response time of the photodiode detectors, which is typically sub-nsec. This is not always the case for photon-counting OCDR, however, since commercially available photon-counting modules typically have far longer response times (≈ several hundred nsec), and therefore saturate at low optical powers. Consequently, collecting an image of a given quality when detector saturation comes into play requires more time when using a photon-counting configuration than when using a conventional configuration. The performance of SSPDs in this respect is superior to that of commercially available single-photon-counting modules, however, as will be discussed in Sec. 5.3.

## 5. Experimental Results

*5.1 Enhancement of axial resolution*

To compare the performance of SSPDs and standard silicon SPADs (single-photon avalanche detectors) in photon-counting coherence-domain reflectometry, an experiment was conducted using the arrangement shown in Fig. 6. A 532-nm (doubled Nd:YVO$_4$) Verdi laser was used to pump a 1.5-mm BBO nonlinear crystal (NLC) cut for type-I phase matching. The crystal was aligned to obtain degenerate and collinear spontaneous parametric downconversion (SPDC). The downconverted light, which served as a convenient broadband optical source centered at 1064 nm, was introduced into a Michelson interferometer. Mirror 1 in the reference arm was placed on a nano-positioning stage to change its position, while mirror 2 was kept stationary. The dichroic components D1, D2, and D3 were used to reflect light at 532 nm and transmit light at 1064 nm; for D1 and D2 the infrared radiation comes from the laser whereas for D3 it comes from the downconversion, which is desired. The Glan–Taylor polarizers P1 and P2 were used to reflect light at 1064 and 532 nm, respectively. The light emerging from P2 was fed into the fiber-coupled detectors (SPAD and SSPD) via a lens.

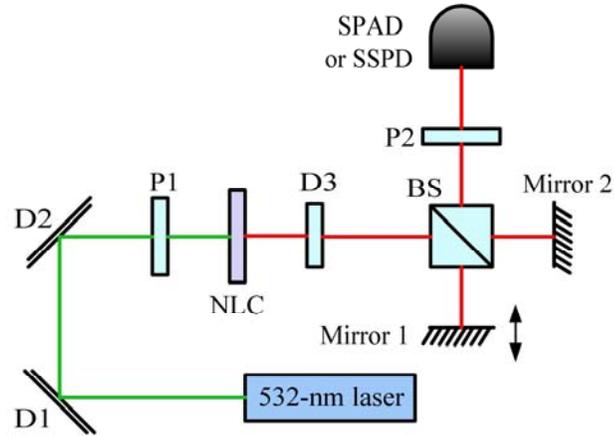

Fig. 6. Photon-counting OCDR experimental arrangement using a Michelson interferometer comprising a beam-splitter (BS) and two mirrors. Mirror 1 is translated to change the length of the reference arm. Collinear spontaneous parametric downconversion generated in a 1.5-mm-thick BBO nonlinear-optical crystal (NLC), cut for type-I phase matching, serves as the optical source. D1 and D2 are dichroic components that direct the 532-nm output of the doubled Nd:YVO$_4$ pump laser to the NLC. Dichroic D3 and Glan–Taylor polarizers P1 and P2 are used to remove unwanted wavelengths. Experiments were performed using both SPADs and SSPDs as photon-counting detectors.

The counts from the SPAD and SSPD were measured in a fixed time window as a function of the position of mirror 1. The resultant interferograms are illustrated in Fig. 7. It is clear from the data that the SSPD offers a narrower interferogram than the SPAD (3.3 vs. 5.4 $\mu$m). In accordance with the discussion in Sections 3.2 and 4.1, this is expected because the SSPD is sensitive over a broader spectral range than the SPAD. This observation, in turn, means that the SSPD offers better axial resolution than the SPAD.

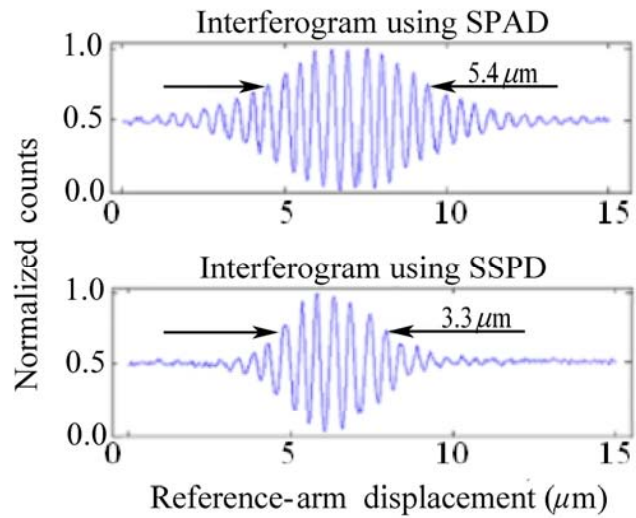

Fig. 7. OCDR interferograms measured with SPAD and SSPD single-photon detectors using the apparatus depicted in Fig. 6. A reduction in the full-width at half maximum (FWHM), corresponding to an improvement in axial resolution, is observed with the SSPD. This is a result of its broader spectral sensitivity.

To better understand the improvement in axial resolution, we calculate the Fourier transforms of the interference signals shown in Fig. 7, and plot them as a function of wavelength. The results, shown in Fig. 8, reveal that the SPAD *is not* sensitive to wavelengths beyond 1100 nm, whereas the SSPD *is* sensitive in this region and therefore yields improved axial resolution. However, the resolution obtained in this experiment is limited by the bandwidth of our downconversion source. Far higher axial resolution could be obtained were we to use an SSPD in conjunction with broader sources that operate near 1100 nm, such as broadband continuum generation from a photonic-crystal fibers [11] and fiber lasers [24], as the SSPD response extends over a far greater wavelength range.

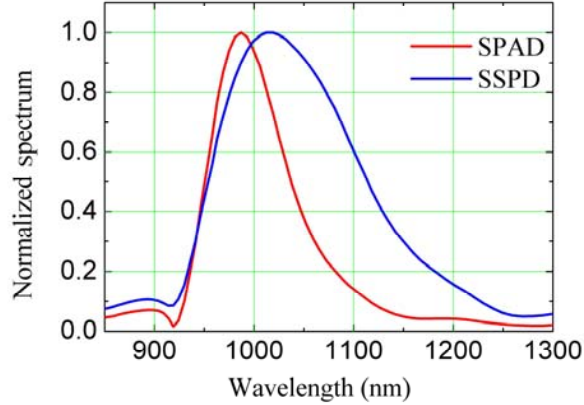

Fig. 8. Fourier transforms of the interference signals shown in Fig. 7, plotted as a function of wavelength. It is evident that the SPAD is not sensitive to wavelengths beyond 1100 nm, whereas the SSPD is sensitive in this region.

*5.2 Enhancement of sensitivity at low light levels*

To demonstrate OCDR using single-photon counting with low levels of source power, we made use of the system depicted in Fig. 2. The source was a standard superluminescent diode (SLD) whose output was centered at a wavelength of 930 nm, with a spectral width of 70 nm. This source, which is often used in coherence-domain imaging, has an optical power that is sufficient so that it can be conveniently measured and attenuated to the level desired for the experiment at hand. The SLD was operated at an output power of ≈ 1 mW, but was attenuated to 10 nW by means of neutral-density (ND) filters placed directly at the output. In addition, to simulate a sample of low reflectance, ND filters were used to introduce an attenuation of 70 dB in the sample arm of the interferometer, which comprised a mirror.

We now forge a comparison with the theoretical results for the SNR provided in Sec. 4.2. The attenuation of 70 dB in the signal arm is expected to result in a signal optical power $P_S \approx$ 2.5 x $10^{-16}$ W (half the power is lost in the interferometer), whereupon $\Phi_S = P_S/h\nu \approx 1170$ photons/sec. Since $\eta$ is measured to be ≈ 0.05 pulses/photon and the effective bandwidth $B$, which is determined by the bandwidth of the digital-filtering system, is ≈ 1/40 Hz (this is narrower than $1/2T$, where $T$ = 1 sec is the counting time per data point). In accordance with Eq. (4), we then expect an SNR ≈ 1170 (30.7 dB). Using the measured envelope of the signal, and the variance of the noise in the region outside the signal (i.e., at a reference-arm displacement greater than the coherence length of the source), we obtain an observed SNR = 562 (27.5 dB), which is within a factor of two of the theoretical prediction.

To examine the count-variance to count-mean ratio, we carried out a series of experiments in which the reference-arm mirror was translated in discrete steps while maintaining the path-length difference between the reference and sample arms within the coherence length of the source ($l_c \approx 6$ $\mu$m). The number of pulses from the detector in 1 sec was measured at each particular location of the reference mirror. A total of $N = 100$ such measurements were made using the SSPD detection system shown in Fig. 2.

A plot of the mean count rate, i.e., the mean number of pulses in a 1-sec counting time, is displayed in Fig. 9(a) as a function of the reference-arm displacement. The error bars denote ±1 standard deviation of the count rate. To confirm whether our observations are in accord with the theory presented in Sec. 4.2 for Poisson statistics, we replot these data in Fig. 9(b) in the form of the observed normalized variance $\hat{F}$. The mean of $\hat{F}$ is indeed seen to be close to unity, and its standard deviation close to $\sqrt{2/N} \approx 0.14$, for all reference-arm displacements. The observation of Poisson counting statistics at different signal magnitudes, corresponding to different reference-arm displacements, indicates that the photon statistics of our source are also Poisson [7]. This demonstrates that the particular SLD used in our experiments is devoid of intensity-fluctuation noise. This, together with the fact that photon counting eliminates thermal noise, is consistent with the use of Eq. (4) for the signal-to-noise ratio.

The results described in this section demonstrate that photon-counting OCDR allows us to achieve nearly shot-noise-limited performance even when using a very weak source of light; this cannot be achieved using conventional detection schemes. It is clear, therefore, that photon-counting coherence-domain imaging can be used to image low-reflectance specimens with a low-power light source.

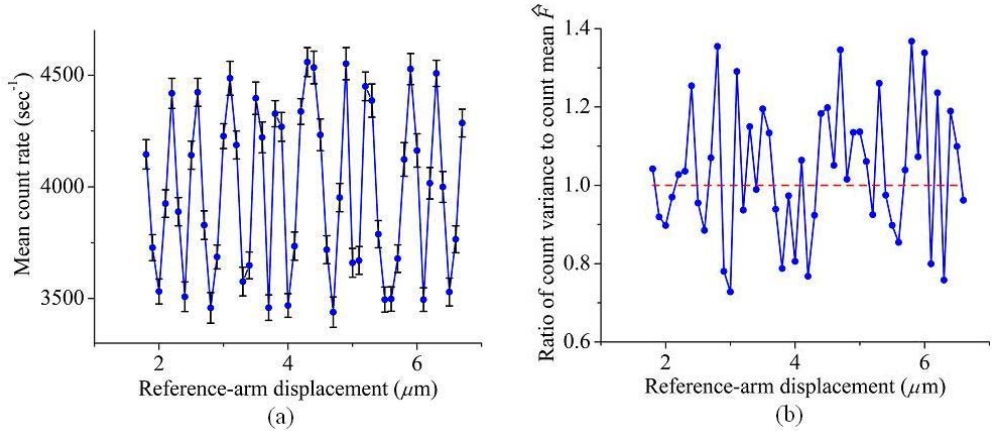

Figure 9. (a) Mean count rate observed at different positions of the reference mirror. Error bars denote ±1 standard deviation. Mean and standard deviations were estimated by taking 100 samples at each mirror position. (b) Ratio of count variance to count mean at different positions of the reference mirror. The measured value of this ratio is close to unity at all positions.

*5.3 Rate of data acquisition*

As indicated in Sec. 4.3, the long response time of single-photon counting detectors limits the rate of data acquisition. However, SSPDs are generally superior to SPADs in this respect. As an example, our SSPDs have a response time of 10 nsec, as shown in Fig. 5.

An experiment was carried out to measure the time required to obtain an OCDR scan of a specified quality. The experimental arrangement is the same as that shown in Fig. 2, using the source described in Sec. 5.2. The SLD was again operated at an output power of $\approx 1$ mW, but in this case ND filters were used to yield a prespecified counting rate. We operated our SSPD at an average rate of 5 MHz corresponding to 50 photons in a counting time of 10 $\mu$sec at the output.

Moving the reference mirror at a speed of 1 mm/sec, scanning for a distance of 1 mm, and using a counting time of 10 $\mu$sec per data point, we observed the two surfaces of a 90-$\mu$m thick silica window, as shown in Fig. 10. We measure a displacement of 134 $\mu$m between the peaks, corresponding to the optical pathlength of $\approx$ 135 $\mu$m, as expected (the refractive index of the silica window is 1.5).

The scan time of 1 sec for the image presented in Fig. 10 could be reduced by a factor of 10 (corresponding to ten times faster scanning of the reference mirror), while maintaining the same image quality, by operating the SSPD at 50 MHz rather than 5 MHz, and using a counting time of 1 $\mu$sec rather than 10 $\mu$sec. Although the SSPD is capable of operating at this rate, we did not use these parameters because of a technical limitation in the speed at which we could move our nanomotion-controlled scanning stage (the maximum speed available was 1 mm/sec). Thus, with a sufficiently fast scanning mechanism, it is evident that SSPDs permit conveniently rapid data acquisition in photon-counting coherence-domain imaging.

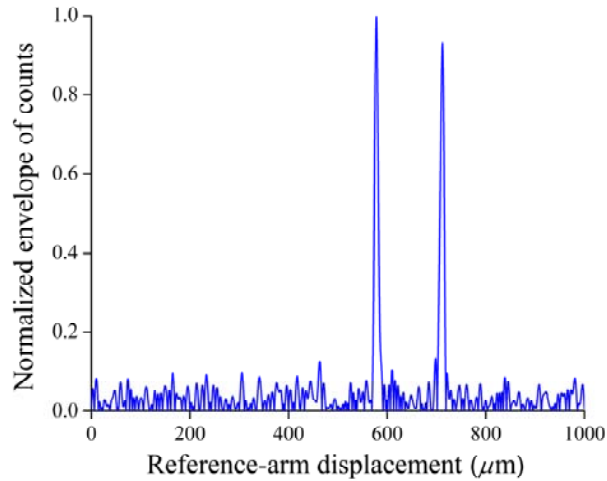

Figure 10: Single-photon axial scan of a 90-$\mu$m-thick silica window obtained with a scanning speed of 1 mm/sec and a counting time of 10 $\mu$sec per data point. The distance between the peaks is 134 $\mu$m, corresponding to the optical pathlength.

## 6. Conclusion

Coherence-domain imaging using single-photon counting allows weak light sources to be used for imaging weakly reflecting samples. We have demonstrated the use of superconducting single-photon detectors (SSPDs) in such an imaging system. These detectors are sensitive over the entire spectral range useful for OCT in biological samples. Neither Si nor InGaAs detectors have comparable sensitivity over the entire spectrum of interest. In addition, SSPDs can also provide high-acquisition-rate imaging, with counting rates as high as 100 MHz, if a sufficient flux of light is available. Although these detectors provide greater flexibility in the choice of optical sources that can be used for coherence-domain imaging, they do require cryogenic cooling, and are more expensive than ordinary semiconductor photodetectors, at least in the current state of our technology.

**Acknowledgments**

This work was supported by a U.S. Army Research Office (ARO) Multidisciplinary University Research Initiative (MURI) Grant; and by the Bernard M. Gordon Center for Subsurface Sensing and Imaging Systems (CenSSIS), an NSF Engineering Research Center. We are grateful to Patrick Callahan for assistance with the software used to run the experiments.